\documentstyle[12pt,aaspp4,flushrt]{article}

\newcommand{\psr}{PSR~J1141$-$6545}
\newcommand{\sm}{M$_{\odot}$}
\def\lapp{\ifmmode\stackrel{<}{_{\sim}}\else$\stackrel{<}{_{\sim}}$\fi}
\def\gapp{\ifmmode\stackrel{>}{_{\sim}}\else$\stackrel{>}{_{\sim}}$\fi}

\begin{document}

\title{Discovery of a Young Radio Pulsar in \\ 
    a Relativistic Binary Orbit}

\author{V. M. Kaspi\altaffilmark{1,2,8},
A. G. Lyne\altaffilmark{3},
R. N. Manchester\altaffilmark{4},
F. Crawford\altaffilmark{1},
F. Camilo\altaffilmark{3,5},
J. F. Bell\altaffilmark{4},
N. D'Amico\altaffilmark{6,7},
I. H. Stairs\altaffilmark{3},
N. P. F. McKay\altaffilmark{3},
D. J. Morris\altaffilmark{3},
A. Possenti\altaffilmark{6}}


\altaffiltext{1}{Department of Physics and Center for Space Research,
Massachusetts Institute of Technology, Cambridge, MA 02139, USA}
\altaffiltext{2}{Department of Physics, Rutherford Physics Building,
McGill University, 3600 University Street, Montreal, Quebec,
H3A 2T8, Canada}
\altaffiltext{3}{University of Manchester, Jodrell Bank Observatory,
Macclesfield, Cheshire SK11 9DL, UK}
\altaffiltext{4}{Australia Telescope National Facility, CSIRO, PO Box
76, Epping, NSW 1710, Australia}
\altaffiltext{5}{Columbia Astrophysics Laboratory, Columbia
University, 550 W. 120th Street, New York, NY 10027, USA}
\altaffiltext{6}{Osservatorio Astronomico di Bologna, via Ranzani,
40127 Bologna, Italy}
\altaffiltext{7}{Istituto di Radioastronomia del CNR, via Gobetti 101, 40129
Bologna, Italy}
\altaffiltext{8}{Alfred P. Sloan Research Fellow}

\begin{abstract}
We report on the discovery of \psr, a radio pulsar in an eccentric,
relativistic 5-hr binary orbit.  The pulsar shows no evidence for
being recycled, having pulse period $P = 394$~ms, characteristic
age $\tau_c = 1.4 \times 10^6$~yr, and inferred surface magnetic
dipole field strength $B = 1.3 \times 10^{12}$~G.  From the mass
function and measured rate of periastron advance, we determine the
total mass in the system to be (2.300 $\pm$ 0.012)~\sm, assuming that
the periastron advance is purely relativistic.  Under
the same assumption, we constrain the pulsar's mass to be $M_p \leq
1.348$~\sm\ and the companion's mass to be $M_c \geq 0.968$~\sm\ (both 99\%
confidence).  Given the total system mass and the distribution of
measured neutron star masses, the companion is probably a massive
white dwarf which formed prior to the birth of the pulsar.  Optical
observations can test this hypothesis.
\end{abstract}


\section{Introduction}

Relativistic binary pulsars have been celebrated as being unique
laboratories for high-precision tests of general relativity and
accurate determinations of neutron star masses
(\cite{tw89,twdw92,sac+98}).  To date, post-Keplerian general
relativistic parameters have been measured for nine binary pulsar
systems (see reviews by \cite{twdw92,tc99,kas99}).  Five are double
neutron star binaries (PSRs B1913+16, B1534+12, B2127+11C, J1518+4904,
J1811$-$1736) and four (PSRs J1713+0747, B1802$-$07, B1855+09,
B2303+46) have white dwarf companions. Through radio timing
observations of PSR~B1913+16, general relativity has been confirmed at
the $\sim$0.3\% level (\cite{tay92}).  Timing observations of these
pulsars show that the neutron star masses fall in a narrow range
centered on 1.35~\sm\ (\cite{tc99}).  The lack of variation of these
masses is surprising given the variety of binary evolution mechanisms
by which they were formed.

Of the above relativistic binary systems, all but one are likely to
share a similar evolutionary history.  The observed pulsar in most of
these systems has a large characteristic age $\tau_c$ and low surface
magnetic field $B$ relative to the bulk of the isolated pulsar
population.  This is because the observed neutron star, the first-born
in the binary, was ``recycled'' by a phase of mass transfer from its
companion as the latter ascended the giant branch (see Bhattacharya \&
van den Heuvel 1991 \nocite{bv91} for details of neutron-star binary
evolution).  This results in spin up of the neutron star and the
reduction of its magnetic field.  The binary observed today has
survived the supernova explosion of the secondary (although in the
cases of PSRs B2127+11C and B1802$-$07, which are in globular clusters,
the possibility of a more complicated evolutionary history
cannot be discounted).  The exception among the above binaries is
PSR~B2303+46 which shows no evidence for having been recycled, having
$B$ and $\tau_c$ comparable to the bulk of the isolated pulsar
population.  PSR~B2303+46, long thought to be part of a double
neutron-star binary, has recently been shown to have a white-dwarf
companion (\cite{vk99}) which must have formed {\it prior} to the
pulsar's birth.  Formation mechanisms for such a system have been
suggested by Portegies Zwart \& Yungelson (1999) and Tauris \& Sennels
(2000) and involve an initial primary, the white dwarf progenitor,
which transferred sufficient matter onto the initial secondary that
the latter underwent core collapse.

We report here on \psr, a relativistic binary pulsar recently
discovered as part of the Parkes Multibeam Pulsar survey.  A
preliminary report on this system was presented by Manchester et
al. (2000).\nocite{mlc+00}

\section{Observations and Results}

\subsection{Discovery Observations}
\label{sec:discovery}

The Parkes multibeam pulsar survey (\cite{lcm+00}), currently underway
at the 64-m radio telescope in Parkes, Australia, makes use of a
multibeam receiver which operates at a wavelength of 20~cm and permits
simultaneous observations of 13 separate regions of the sky
(\cite{swb+96}).  The survey, roughly halfway complete, has been
extremely successful, having discovered over 500 new radio pulsars
(\cite{clm+00,dlm+00}).  The observing system used has been described
by Lyne et al. (2000). \nocite{lcm+00} Observations are centered on
1374~MHz, with 288~MHz of bandwidth in each of two orthogonal
polarizations, for each of the 13 beams.  The filtered and amplified
signals are fed into a $2 \times 13 \times 96 \times 3$~MHz filterbank
system, then are square-law detected.  The polarization pairs are
summed, and the results are 1-bit digitized every 250~$\mu$s and
recorded on tape.  Observations are of 35-min duration.  Offline, the
data are dedispersed at many trial dispersion measures (DMs) and
subjected to $2^{23}$-point Fast Fourier Transforms.

\psr\ was discovered in our standard analysis.  A periodicity at
394~ms and at DM of 118~pc~cm$^{-3}$ was reported as having
signal-to-noise ratio 15.4 in data obtained on 1999 April 28.
Immediately apparent in the discovery observation was the variation in
the pulse period due to binary acceleration.  Had the pulsar been
isolated, its signal-to-noise ratio would have been 40.  Confirmation
observations were made at Parkes on 1999 July 10.  A series of
observations in the following week determined the binary parameters.

\subsection{Timing Observations}
\label{sec:timing}

Regular timing observations of \psr\ were carried out at numerous
epochs from 1999 July 10 through 2000 January 19 using the central
beam of the multibeam receiver.  In the timing observations, the same
data acquisition and online software system are used as in the regular
survey observations.  At 1374~MHz, we obtained 123 pulse arrival times
at 3-MHz frequency resolution, and 31 pulse arrival times using a higher
resolution $2 \times 512 \times 0.5$-MHz filterbank.  Two observations
were made at 600~MHz, on 1999 August 23 and 24, using a $2\times 256
\times 0.125$-MHz filterbank and the standard data acquisition system.

Timing observations were divided into 1-min sub-integrations which
were partially de-dispersed into 8 sub-bands and folded at the
predicted topocentric pulsar period using 1024 pulse phase bins.
Typical total integration times were $\sim$10~min.  Final dedispersion
of the sub-bands and folding of the sub-integrations were done in a
subsequent processing step.  Each pulse profile was then
cross-correlated with a high signal-to-noise template appropriate for
the observing frequency and filterbank resolution.  The template for
the high-resolution 20-cm observations is shown in
Figure~\ref{fig:grand}.  The other templates appear similar, including
that at 600~MHz.

The pulse times-of-arrival (TOAs) which result from the
cross-correlation were then subject to a timing analysis using the
{\tt TEMPO} software package\footnote{see
http://pulsar.princeton.edu/tempo} which converts them to barycentric
TOAs at infinite frequency using the Jet Propulsion Laboratory DE200
solar-system ephemeris (\cite{sta82}) and performs a multi-parameter
least-squares fit for the pulsar spin and binary parameters.  During
the fitting, the pulsar's position was held fixed at the value
determined by interferometric observations (see \S\ref{sec:atca}).
Also, the DM was determined using the two 600-MHz observations and six
1400-MHz observations obtained on 1999 August 22 and 23, and was held
fixed for the rest of the timing analysis.

\begin{figure}
\plotone{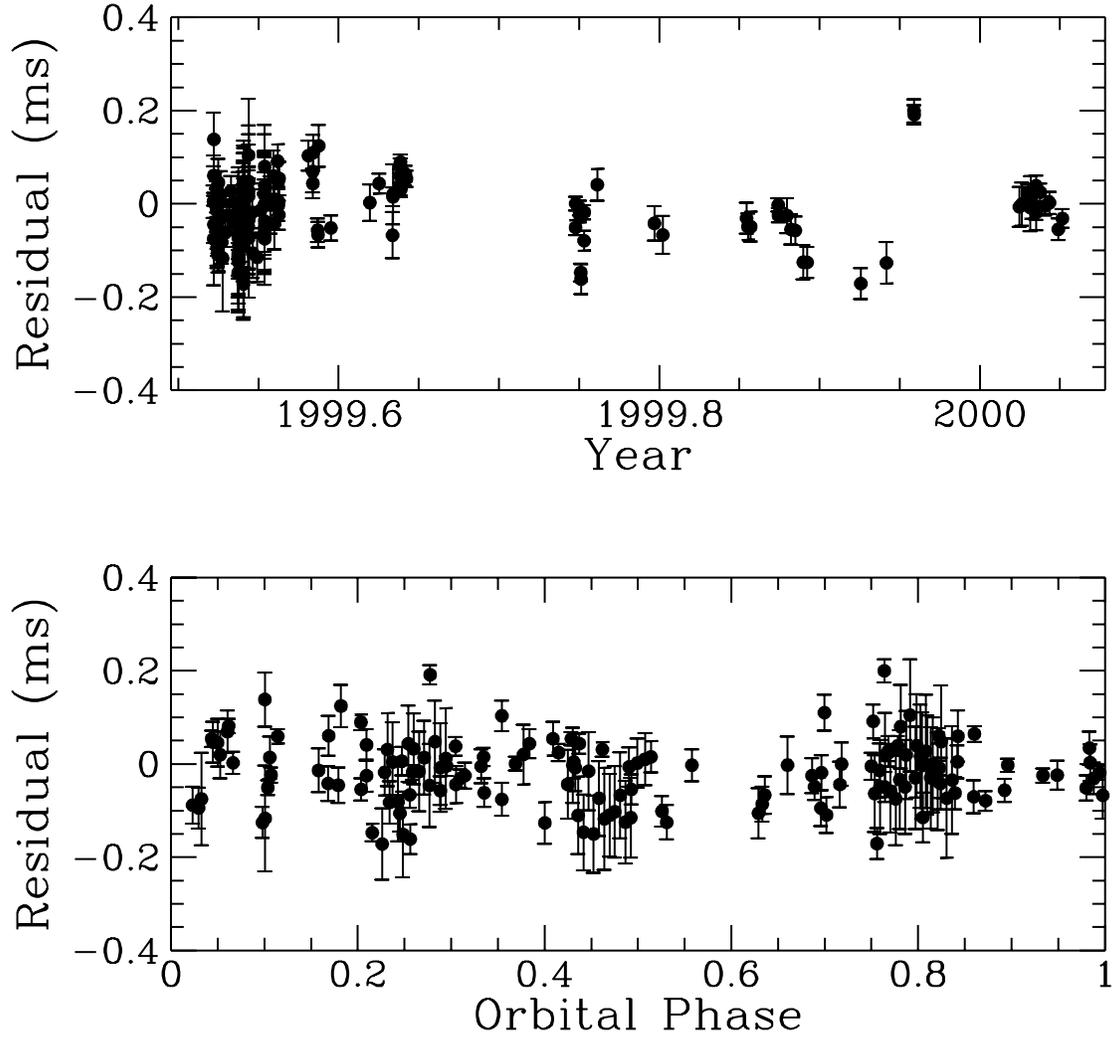}
\figcaption[fig2.eps]{Timing residuals for \psr.  The top panel shows
residuals as a function of time, the bottom panel as a function of
orbital phase, where phase 0/1 is periastron.  The RMS timing residual 
is 66~$\mu$s ($1.7 \times
10^{-4})P$. \label{fig:res}}
\end{figure}

The results of the timing analysis are presented in
Table~\ref{ta:parms}.  The fit parameters were $P$, $\dot{P}$, the
five Keplerian orbital parameters, and a rate of periastron advance,
$\dot{\omega}$.  Upper limits on other post-Keplerian parameters were
determined individually.  Post-fit timing residuals are shown in
Figure~\ref{fig:res}.  The uncertainties in the arrival times were
determined empirically by demanding that the reduced $\chi^2$ for
observations over a short span (1--2 days) be unity.  The formal
reduced $\chi^2$ for these residuals is 4.7 (145 degrees of freedom).
This clearly indicates that there are unmodeled effects in the data.
That the residuals as a function of orbital phase do not display
systematic trends leads us to believe the orbital parameters are not
biased by these effects.  Indeed the two arrival times with the
largest residuals, near epoch 1999.96, are at orbital phases separated
by $\sim$0.5.  The unmodeled features may be manifestations of
standard ``timing noise,'' a stochastic process seen in many young
pulsars like \psr\ (e.g. \cite{antt94}).  This may bias the measured spin
parameters slightly.

\subsection{Radio Interferometric Observations}
\label{sec:atca}

\psr\ was observed on 1999 August 22 and 23 with the Australia
Telescope Compact Array (ATCA) which is located in Narrabri,
Australia.  The array was in its ``6D'' configuration, which has
maximum baseline 5878~m.  The observations were done in pulsar gating
mode at center frequencies of 1384 and 2496~MHz simultaneously, with
128~MHz of bandwidth in each of two linear polarizations for each
frequency.  The primary and secondary calibrators for the observations
were the bright radio point sources 1934$-$638 and 1329$-$665.  The
data were processed using the {\tt MIRIAD} processing
package.\footnote{http://www.atnf.csiro.au/computing/software/miriad/}
After standard interference flagging and calibration were done, on-
and off-pulse images were formed.  The off-pulse emission was then
subtracted from the UV data set leaving only on-pulse emission from
the pulsar.  A point source was then fitted to these UV data using the
nominal position and a rough estimate of the flux density of the
pulsar.  A model of the resulting UV data was then fitted using trial
positions and flux densities for the pulsar using the {\tt MIRIAD
UVFIT} task. Best-fit values and uncertainties were obtained for both
the 1384- and 2496-MHz data sets.  Since the pulsar is brighter at
1384~MHz, we use the more accurate fitted position from that data set
in the timing analysis (see Table~\ref{ta:parms}).  The 1384-MHz
pulsed flux ($\sim$3.3~mJy) is consistent with that estimated from the
Parkes observations; the $3\sigma$ upper limit on point-like radio
emission off-pulse is 0.66 mJy at this frequency.
 
\subsection{Polarization}
\label{sec:poln}

\begin{figure}
\plotfiddle{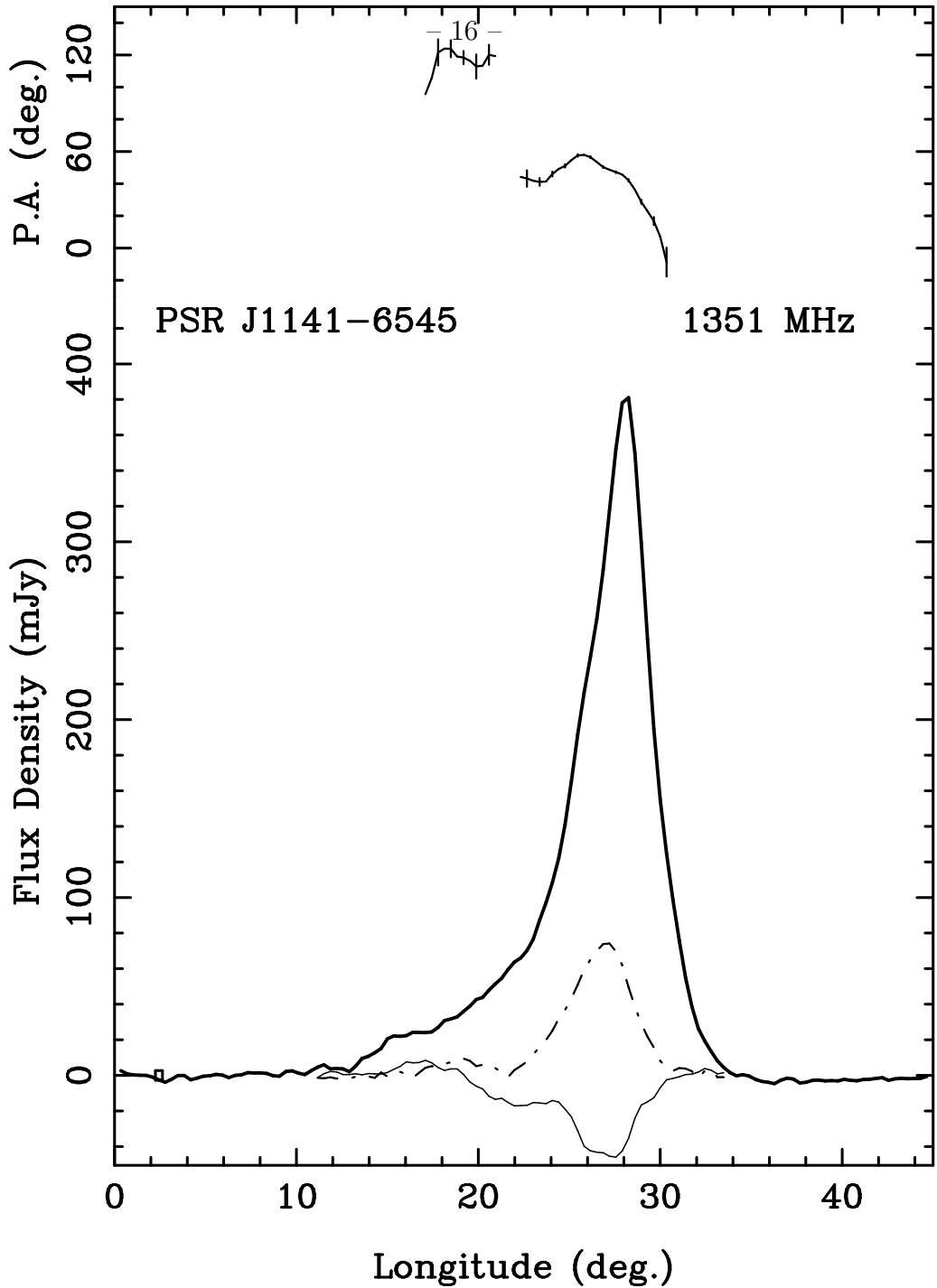}{7in}{0}{80.0}{80.0}{-220}{0}
\figcaption[fig3.eps]{Polarization parameters for PSR J1141$-$6545 at 1351 MHz. In the
lower part of the figure, the pulse total intensity (Stokes $I$) is given by
the thick line, the linearly polarized intensity ($L = (Q^2 + U^2)^{1/2}$)
by the dot-dash line and the circularly polarized intensity (Stokes $V$) by
the thin line. $V$ is defined in the sense that LH-circular (IEEE
convention) is positive. The upper part of the figure gives the position
angle above the Earth's ionosphere, measured from north toward east, of the
linearly polarized component. \label{fig:poln}}
\end{figure}

We have measured the polarization properties of \psr, which are
especially interesting given the possibility of significant precession
of the pulsar (see \S\ref{sec:relobs} below).  Observations were made
on 2000 March 26 (MJD 51629) using the center beam of the multibeam
receiver and the Caltech correlator (Navarro
1994)\nocite{nav94}. Instrumental phases and gains were calibrated
using a pulsed noise signal injected at $45^{\circ}$ to the two signal
probes (see Navarro et al. 1997\nocite{nms+97} for a full
description). Two 25-min observations using 64-MHz bandwidth at
orthogonal feed angles were summed to produce the polarization
profiles shown in Figure~\ref{fig:poln}.  Position angles were
corrected for Faraday rotation in the Earth's ionosphere. The
ionospheric contribution to the rotation measure (RM) was estimated to
be $-3.2$ rad m$^{-2}$ at the time of the observation using an
ionospheric model, so the plotted angles are approximately
$9^{\circ}$ greater than those observed. The precision of the absolute
position angles is limited by the uncertainty in this correction,
which we estimate to be $\pm 2^{\circ}$.

On average, the pulse profile is 16\% linearly polarized (see
Figure~\ref{fig:poln}).  The position angle shows a complicated
variation across the pulse with a clear orthogonal transition in the
leading wing and a sharp drop toward the trailing edge of the pulse.
The circular polarization is relatively strong, with a mean value of
$-10$\%. There is a possible reversal in the sign of $V$ at a pulse
phase slightly preceding the orthogonal transition in the linear
polarization. These properties are not consistent with any simple
model for the polarization. The shape of the total intensity profile
suggests that the main peak is near the trailing edge of a conal
beam. This may be supported by the location of the reversal of sign of
$V$ which is usually (but not always) located near the center of the
beam (Han et al. 1998)\nocite{hmxq98}. However, the observed position
angle variation is certainly not consistent with a rotating vector
model having a beam center on the leading wing.

The interstellar component of the RM, estimated by computing the
weighted mean position angle difference between the two halves of the
observed band, is $-86 \pm 3$ rad m$^{-2}$. This implies a mean
line-of-sight interstellar field, weighted by the electron density
along the path, of 0.9 $\mu$G directed away from us.  RMs of
neighboring pulsars are of mixed sign and magnitude (Han, Manchester
and Qiao 1999)\nocite{hmq99}. At the estimated distance of 3.2 kpc, as
inferred from the DM and a model for the Galactic electron
distribution (\cite{tc93}), the path to the pulsar crosses the Carina
arm; the structure of the Galactic magnetic field in this region is
evidently complex.

\section{Discussion} \label{sec:disc}

\psr\ is different from most of the other known relativistic binary
pulsars in several important respects.  The pulsar's characteristic
age, $\tau_c = 1.4 \times 10^{6}$~yr, and inferred surface magnetic
field strength, $B = 1.3 \times 10^{12}$~G, are similar to those of
the bulk of the isolated pulsar population. Thus, \psr\ is unlikely to
have ever been recycled.  Like the other short-$P_b$, eccentric
relativistic binaries, it may be a double neutron star binary, but if
so, we must be observing the second-formed neutron star in its
short-lived radio pulsar phase.  In this case, its companion could
still be an observable radio pulsar that was not detected in our
Parkes observations, either because its short spin period varies too
rapidly due to the binary orbit (a possibility we are checking, given
our knowledge of the binary orbit) or because its radio beam does not
intersect our line of sight.

However, we show here that the companion is unlikely to be a second
neutron star, given the system's mass function, its rate of precession
$\dot{\omega}$, and what is known about the neutron star mass
distribution.   The mass function is given by
\begin{equation}
f(M_p) = \frac{4 \pi^2 (a \sin i)^3}{G P_b^2} =  \frac{M_c^3 \sin^3
i}{(M_p + M_c)^2},
\end{equation}
where $M_p$ and $M_c$ are the pulsar and companion masses, respectively,
$P_b$ is the orbital period, and $a \sin i $ is the projected semi-major
axis of the pulsar orbit.  This provides a constraint on the 
minimum $M_c$ for any assumed
$M_p$, by setting $i$, the inclination of the orbital angular momentum with
respect to the line-of-sight, to $90^{\circ}$.  In addition, $\dot{\omega}$,
under the assumption that it is due to general relativistic
periastron precession (but see \S\ref{sec:classom}), is given by
\begin{equation}
\dot{\omega}_{GR} = 3 \left( \frac{P_b}{2 \pi} \right)^{-5/3} (T_{\odot}
M)^{2/3} (1 - e^2)^{-1},
\end{equation}
where $T_{\odot} = 4.925490947 \times 10^{-6}$~s.  This then
determines the total system mass $M = M_p + M_c$.  For $\dot{\omega} =
(5.32 \pm 0.02)^{\circ}$yr$^{-1}$
(Table~\ref{ta:parms}), $M = (2.300 \pm 0.012)$~\sm.
Figure~\ref{fig:mass} shows these constraints in $M_p - M_c$ phase
space.  The shaded region is ruled out by the mass function.  The
intersection of the $\dot{\omega}$ straight line with the boundary of
this region
determines the maximum allowed pulsar mass: $M_p < 1.331$~\sm\ ($1\sigma$)
or $M_p < 1.348$~\sm\ ($3\sigma$).  Similarly, we set a lower limit
$M_c > 0.974$~\sm\ ($1\sigma$) or $M_c > 0.968$~\sm\ ($3\sigma$).

Thorsett \& Chakrabarty (1999) \nocite{tc99} showed that measurements
of neutron star masses are consistent with all being drawn from a
Gaussian distribution having mean and standard deviation 1.35 and
0.04~\sm, respectively.  The parameters in the \psr\ system are thus
interesting: if $M_p$ is close to its maximum allowed value
corresponding to $i \simeq 90^{\circ}$, as would be consistent with
other neutron star masses, then its companion has mass much lower than
those of all known neutron stars, and therefore is unlikely to be one.
Of course on {\it a posteriori} statistical grounds, $i \simeq
90^{\circ}$ is improbable.  On the other hand, a smaller $i$ implies a
less massive pulsar.  For the median value $i = 60^{\circ}$, $M_p =
1.17$~\sm, and $M_c = 1.13$~\sm, both significantly lower than for any
other known neutron stars.  Note that none of these values can be
ruled out based on neutron star stability arguments, since masses of
as little as $\sim$0.1~\sm\ are allowed (e.g. \cite{cst93}).  However
the formation mechanism for such low-mass neutron stars is unclear.

\subsection{A Neutron Star/White Dwarf Binary?}

A more likely possibility is that the \psr\ system is a neutron
star/CO white-dwarf binary seen edge-on.  (ONeMg white dwarfs have
minimum mass $\sim$1.1~\sm [Wanajo, Hashimoto \& Nomoto 1999]
\nocite{whn99} so this possibility cannot be ruled out.)  The
evolutionary history of the \psr\ binary, if the companion is a
massive white dwarf, is then clear (see e.g. \cite{dc87}, \cite{py99},
\cite{ts00}): the system originated as a binary consisting of two main
sequence stars having mass ratio near unity, but with neither
sufficiently massive to independently form a neutron star.  For
example, the primary may have had mass 7~\sm\ and the secondary 5~\sm.
The primary evolved first to form the white dwarf, in the process
transferring sufficient matter onto the secondary for its mass to
exceed that necessary to form a neutron star.  After the white dwarf
formed, it spiraled into the envelope of the now more massive
secondary as the latter ascended the giant branch, greatly decreasing
the orbital period, and ejecting the common envelope.  The secondary,
a helium star after the ejection of the envelope, then exploded in a
supernova, which fortuitously did not disrupt the binary.  The radio
pulsar we see is therefore the second evolved star.

The evolutionary scenario in which a massive white dwarf forms prior
to the neutron star has been considered in population synthesis
studies.  Dewey \& Cordes (1987) \nocite{dc87} and Portegies Zwart \&
Yungelson (1999) \nocite{py99} showed that the birth rate of white
dwarf/young pulsar binaries is comparable to that of double neutron
star binaries.  However, Portegies Zwart \& Yungelson (1999) argued
that the white dwarf should have mass $\gapp 1.1$~\sm, a result of
their assumption that a neutron-star forming binary had to have an
initial primary mass of $>7$~\sm.  To form a $\sim$1~\sm\ white dwarf,
an initial minimum primary mass of $\sim$6~\sm\ is required
(S. Portegies Zwart, private communication).  The work of Tauris \&
Sennels (2000) \nocite{ts00} confirms that white dwarf/young neutron
star systems should be observable, but they suggest that the birth
rate of such systems is much higher than that of double neutron star
systems.  A lower allowed initial primary mass for forming white
dwarfs in the Portegies Zwart \& Yungelson (1999) analysis could
reduce the disagreement.

One prediction of the binary evolution theory is that the space
velocity of \psr\ will be greater than $\sim$150~km~s$^{-1}$
(\cite{ts00}).  At a distance of 3.2~kpc, this implies a proper
motion of $> 10$~mas~yr$^{-1}$, which could be measurable by timing
on a time scale of a few years.  VLBI observations may also be able to
detect it.  In fact, given the pulsar's timing age, Galactic latitude,
distance estimate, and assuming that the pulsar was born in the Galactic
plane, the component of the pulsar's space velocity perpendicular to
the plane must be $\sim150(d/3.2 \; {\rm kpc})$~km~s$^{-1}$, already
roughly consistent with the prediction.


If the companion is indeed a white dwarf, allowing for interstellar
reddening, its B magnitude is likely to be in the range 25 -- 26.5 and
R in the range 24.5 -- 26. From a Digital Sky Survey image of the
field, it is clear that the field is relatively crowded, but not
impossibly so. Excellent seeing conditions on the VLT would allow
detection of the companion. If detected, then a follow-up temperature
measurement using HST may provide a useful age constraint.

\subsection{Classical Contribution to $\dot{\omega}$?}
\label{sec:classom}

The observed $\dot{\omega}$ may not be purely relativistic, in which
case the above conclusions would have to be modified.  Classical
contributions to $\dot{\omega}$ can come either from tidal deformation
of the companion by the neutron star (significant only if the
companion is non-degenerate), or from a rotation-induced quadrupole
moment in the companion, possibly relevant if it is a
rapidly rotating white dwarf.  We consider these possibilities in
turn.

\begin{figure}
\plotone{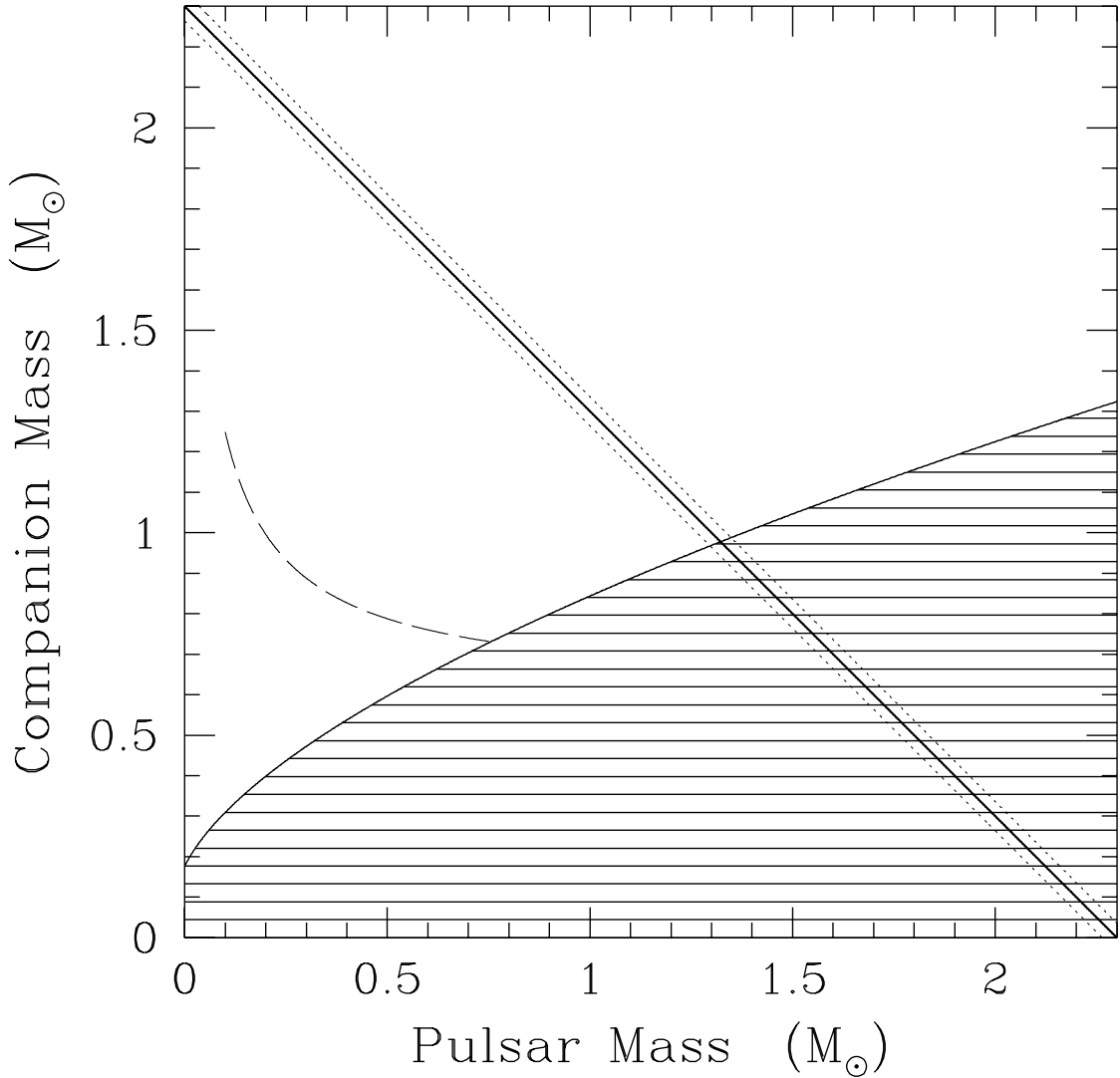} 
\figcaption[fig4.eps]{Allowed companion mass
($M_{\rm c}$) and pulsar mass ($M_{\rm p}$) phase space.  The shaded
area is ruled out by the mass function.  The solid thick straight line
shows the constraint on the total mass $M$ implied by $\dot{\omega}$,
assuming it is purely relativistic.  The dotted lines representing the
3$\sigma$ uncertainties. The dashed line shows allowed masses for a
non- or slowly rotating helium star companion, under the assumption
that the observed $\dot{\omega}$ is partially due to a tide raised by
the neutron star (see \S\protect\ref{sec:classom}). \label{fig:mass}}
\end{figure}

A tide raised on the companion star gives it a quadrupole moment
that results in a classical
apsidal advance, $\dot{\omega}_{tide}$, given by
\begin{equation}
\dot{\omega}_{tide} = 3.44 \times 10^6 k_2 \left(\frac{M_p}{M_c}\right)
\left( \frac{M}{M_{\odot}}\right)^{-5/3} \left( \frac{R_c}{R_{\odot}}
\right)^5 \;\; 
  ^{\circ}{\rm yr}^{-1},
\end{equation}
where $R_c$ is the companion radius and $k_2$ its apsidal constant, a
measure of its internal density distribution (\cite{rma76,sb76}).  A
hydrogen main sequence star companion could in principle fit in the
orbit, but would result in a classical $\dot{\omega}$ $\sim$1000 times
the relativistic value (see \cite{mr75}) so is certainly ruled out.
However, a helium
main sequence star has much smaller radius, and would result in a
classical $\dot{\omega}$ that is comparable to the expected
relativistic value.  We have explored this possibility by considering
a range of values for $M_p$ and $M_c$, and finding the expected
$\dot{\omega} = \dot{\omega}_{tide} + \dot{\omega}_{GR}$ that match
the observed value.  For these calculations, we assumed
$k_2(R_c/R_{\odot})^5 = 4 \times 10^{-6}(M_c/{\rm M}_{\odot})^{4.59}$, as
derived by Roberts et al. (1976).  In Figure~\ref{fig:mass}, the
dashed line shows the allowed locus in $M_p - M_c$ space for a non- or
slowly rotating 
helium star companion.  Clearly this is possible only if
$M_p \leq 0.75$~\sm.  This seems unlikely given observed
masses of other neutron stars.

A rapidly rotating white dwarf would have a quadrupole moment that
could result in a significant classical $\dot{\omega}$ (Roberts et
al. 1976, \cite{sb76}).  However unlike the tidal quadrupole, the
rotationally-induced quadrupole lies in a plane perpendicular to the
white dwarf spin axis.  Since the pulsar formed after the white dwarf,
it is unlikely that the white dwarf spin axis is aligned with the
orbital angular momentum, as this would demand a fortuitously
symmetric supernova explosion.  A misalignment of the angular momenta
results in classical spin-orbit coupling, that is, a precession of the
orbital plane, $di/dt$ as in the pulsar/B-star binary PSR~J0045$-$7319
(\cite{lbk95,kbm+96}).  This precession would manifest itself as a
time-variable projected semi-major axis $x \equiv a \sin i$.  In
general, the magnitude of $di/dt$ is comparable to that of the
periastron precession $\dot{\omega}$ (\cite{wex98}).  For \psr, we
find an upper limit on $\dot{x}$ that is much smaller than
$\dot{\omega}$ (Table~\ref{ta:parms}).  This might suggest that
spin-orbit coupling is unlikely to be occurring.  However, $\dot{x}$
varies as $\cos i$; since $i \simeq 90^{\circ}$ (especially if
$\dot{\omega}_{GR}$ is smaller than our observed value), we cannot
rule out orbital plane precession, i.e. a large $di/dt$.

The above reasoning also holds for a helium-star companion that is
rotating rapidly.  In this case, the dashed line in
Figure~\ref{fig:mass} is inappropriate as it assumes no contribution
from a rotation-induced quadrupole moment.  Hence, we cannot presently
rule out this possibility.  However, a 1-\sm\ helium main sequence
star would be considerably brighter than a white dwarf: assuming it is
on the helium main sequence, it would have bolometric luminosity $\log
(L/L_{\odot}) = 2.4$ and $T_{eff} = 50,000$~K (\cite{kw90a}).  Using
bolometric corrections from Bessell, Castelli \& Plez (1998)
\nocite{bcp98} and given the distance and expected reddening, a helium
star should have $B \simeq 17$, and should be easily distinguishable
from a white dwarf.

\subsection{Future Relativistic Observables}
\label{sec:relobs}

The prospects for a high-precision determination of both component
masses in the \psr\ binary system are excellent.  Our upper limit on the
combined time dilation and gravitational redshift parameter $\gamma$
is shown in Table~\ref{ta:parms}.  Note that this is largely independent
of which model for the system is correct.  Given that the fractional
uncertainty in $\gamma$ scales as $T^{-3/2}$, where $T$ is the
observing span (\cite{dt92}), we expect a 3$\sigma$ detection by 2001,
assuming $\dot{\omega}$ is purely relativistic (for which $\gamma
\simeq 0.7$~ms for $M_p = 1.32$~\sm).  Furthermore, a general
relativistic orbital period derivative, $\dot{P_b}$, should be
measurable with interesting precision by 2004, given that its
fractional uncertainty scales as $T^{-5/2}$ and its expected value is
$-3.5 \times 10^{-13}$.  Thus, continued high-precision timing
observations of \psr\ will eventually test our assumptions about a
purely relativistic $\dot{\omega}$.  Although the orbit is likely to be
highly inclined, measurement of Shapiro delay will be difficult: the
maximum expected delay is only $\sim$5~$\mu$s, well below our current
timing precision.

Although \psr\ will coalesce in $\sim$1.5~Gyr
due to gravitational-radiation-induced orbit
decay, it, and coalescing neutron-star/white-dwarf binaries in
general, will not be detected by LIGO.  The detectability depends
on the frequency of the emitted radiation near the time of
coalescence.  This depends on the size of the orbit at coalescence,
which, to order of magnitude, is the white dwarf radius, $\sim
10^9$~cm.  This implies an orbital frequency of $\sim$0.25~Hz at
coalescence; the implied gravitational wave frequency, twice the
orbital frequency, is well outside the LIGO band of 10--10,000~Hz.
However, such signals would be within the 0.00001--1~Hz band of the
proposed Laser Interferometric Space Antenna (LISA; \cite{ben98}).

\begin{figure}
\plotone{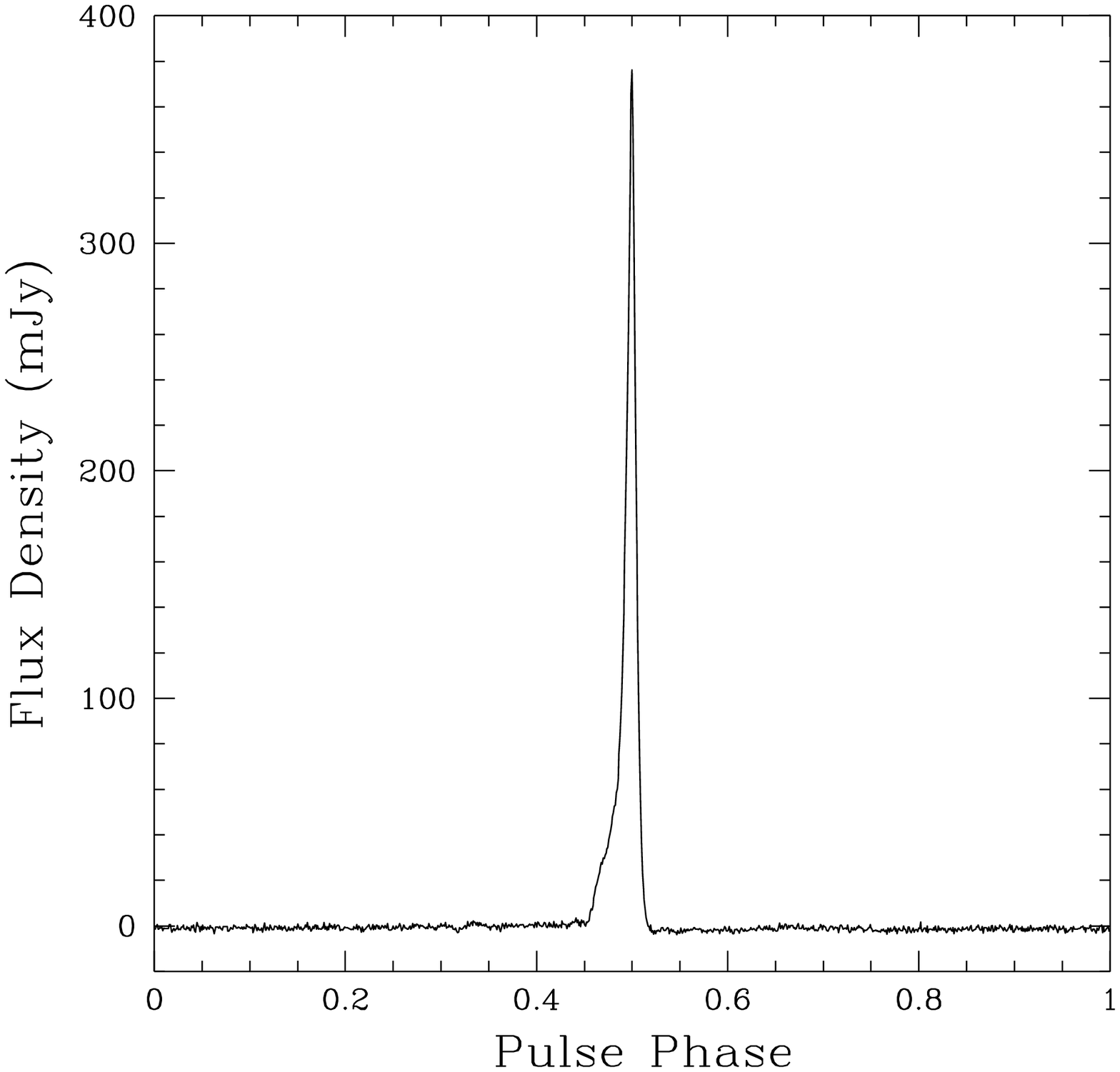} 
\figcaption[fig1.eps]{Average pulse profile for
\psr\ at 20~cm, obtained with 500~kHz frequency resolution.  The pulse
period is 394~ms.  The effective dispersion smearing is 0.19~ms or
$(4.8 \times 10^{-4})P$. \label{fig:grand}}
\end{figure}

General relativistic geodetic precession (\cite{bo75}) is predicted to
result in the precession of the pulsar's spin axis at a rate of $\sim
1.7^{\circ}$yr$^{-1}$ (assuming a purely relativistic $\dot{\omega}$).
This is greater than that predicted for the two pulsars for which it
has been observed, PSR~B1913+16 (\cite{wrt89}; \cite{kra98}) and PSR~B1534+12
(\cite{arz95,stta00}), although the measured amplitude will be
suppressed by a possibly large factor that depends on the unknown
line-of-sight geometry and spin-orbit misalignment.  Given the narrow
pulse profile seen in \psr\ (Fig.~\ref{fig:grand}, FWHM of 0.011$P$),
this effect could be detected in a few years, barring unfortunate
geometries.  Thus far, we have detected no strong evidence for pulse
profile changes, although a detailed analysis is beyond the scope of
this paper.

That the pulsar was not detected in past pulsar radio surveys of the
region is intriguing.  The Johnston et al. (1992) \nocite{jlm+92}
20-cm survey of the Galactic plane had minimum detectable flux density
of $\sim$1~mJy, below the pulsar's 20-cm flux density of 3.3~mJy.  At
this frequency, the pulsar displays no evidence for scintillation. The
Parkes all-sky survey at 70~cm may also have been able to detect it.
That survey's sensitivity reached 20~mJy for Galactic plane sources;
extrapolating the flux densities in Table~\ref{ta:parms}, the
estimated flux density at 70~cm is $\sim$40~mJy. Scattering for the
pulsar at that frequency is negligible. Thus, it is plausible that the
pulsar's beam has only recently precessed into our line-of-sight, a
result of geodetic precession.  If so, the pulse profile should be
evolving rapidly.

\bigskip

We thank M. Kramer for assistance at Parkes, B. Gaensler
for help with the ATCA data, and V. Kalogera, M. van
Kerkwijk, S. Portegies Zwart, F. Rasio and S. Thorsett for helpful 
discussions.  VMK is supported by a National Science Foundation CAREER
award (AST-9875897).  FC is supported by NASA grant NAG~5-3229.



\clearpage

\clearpage

\clearpage

\clearpage

\clearpage
\begin{deluxetable}{lc}
\tablecaption{Measured and Derived Parameters for \psr. \label{ta:parms}}
\tablewidth{0pt}
\tablehead{
\colhead{Parameter} & \colhead{Value} }
\startdata
Right Ascension\tablenotemark{a,b} $\;\;$ (J2000) & 11$^{\rm h}$ 41$^{\rm
m}$ 07$^{\rm s}$.053(15) \\
Declination\tablenotemark{a,b} $\;\;$ (J2000) & $-$65$^{\circ}$
45$'$ 18$''$.85(10) \\
Galactic Longitude\tablenotemark{a,b}  & $295^{\circ}.79$ \\
Galactic Latitude\tablenotemark{a,b}   & $-3^{\circ}.86$ \\
Dispersion Measure\tablenotemark{b}, $\;$ DM (pc cm$^{-3}$) &
116.017(11)\\
Flux Density\tablenotemark{a} $\;$ at 1384 MHz (mJy) & 3.3(5) \\
Flux Density\tablenotemark{a} $\;$ at 2496 MHz (mJy) & 0.9(5) \\
Pulse FWHM at 20~cm & 0.011$P$ \\
Rotation Measure, RM (rad m$^{-2}$) & $-$86(3)\\\tableline
Period, $P$ (s) & 0.3938978339002(22)\\ 
Period Derivative, $\dot{P}$ & 4.3070(2)$\times 10^{-15}$\\
Epoch (MJD\tablenotemark{c}~) & 51369.8525 \\
Orbital Period, $P_b$ (days) & 0.197650965(5) \\
Eccentricity, $e$ & 0.171881(9) \\
Projected Semi-Major Axis, $a\sin i \equiv x$ (light s) & 1.85945(1) \\
Longitude of Periastron, $\omega$  & 42$^{\circ}$.436(5) \\
Epoch of Periastron, $T_0$ (MJD\tablenotemark{c}~) & 51369.854549(3) \\
Advance of Periastron, $\dot{\omega}$ ($^{\circ}$yr$^{-1}$) & 5.33(2) \\
Relativistic Time Dilation and Gravitational Redshift, $\gamma$ (ms) &
$<2.5$ (3$\sigma$) \\
Orbital Period Derivative, $|\dot{P_b}|$ & $<5 \times 10^{-11}$
(3$\sigma$) \\
Rate of Change of $a\sin i$, $|\dot{x}|$ (light s s$^{-1}$) & 
$<5 \times 10^{-12}$ (3$\sigma$) \\\tableline
Characteristic Age, $\tau_c$ (Myr) & 1.4 \\
Inferred Surface Dipole Magnetic Field Strength, $B$ (G) & $1.3 \times 10^{12}$\\
Spin-Down Luminosity, $\dot{E}$ (erg s$^{-1}$) & $6.9 \times  10^{32}$ \\
Mass Function, (\sm) & 0.176701(3) \\
Total System Mass, $M$ (\sm) & 2.300(12)\\
\enddata

\tablenotetext{a}{Parameter determined from ATCA observations.  See
\S\ref{sec:atca}.}
\tablenotetext{b}{Parameter determined at a single epoch and held fixed in
timing analysis.  See \S\ref{sec:timing}.}
\tablenotetext{c}{Modified Julian Day MJD = JD - 2400000.5.}

\tablecomments{Numbers in parentheses represent 1$\sigma$ uncertainties
in the last digit quoted.}

\end{deluxetable}

\end{document}